\documentclass[noshowpacs,amsmath,
twocolumn,
superscriptaddress,
8pt
]{revtex4-1}
\usepackage{setspace}
\usepackage{amsmath}
\usepackage{float}
\usepackage{bm}
\usepackage{graphicx}
\usepackage[nearskip,margin = 0pt]{subfig}

\usepackage{verbatim}
\usepackage{amsfonts}
\usepackage{braket}
\usepackage{amssymb}
\usepackage{upgreek}
\usepackage[colorlinks,linkcolor=cyan,anchorcolor=blue,citecolor=cyan,urlcolor=black]{hyperref}
\usepackage{epstopdf}
\usepackage{xcolor}
\usepackage{booktabs}
\usepackage{tabularx}
\usepackage{xtab}
\usepackage{changepage}
\usepackage{ragged2e}

\DeclareGraphicsExtensions{.pdf,.eps,.png,.jpg,.mps}

\begin{document}
\title{Defect-Assisted Domain Nucleation Drives Unique Exchange Bias Phenomena in $\bf{MnBi_2Te_4}$}



\author{Shiqi Yang,$^{1,\dagger}$ Xiaolong Xu,$^{2,\dagger}$ Yuchen Gao,$^{1,{\dag}}$ Roger Guzman,$^{3}$ Pingfan Gu,$^{1}$ Huan Wang,$^{4}$ Yuan Huang,$^{2,\star}$ Wu Zhou,$^{3,\star}$ Tianlong Xia,$^{4,\star}$ and Yu Ye$^{1,5,6,7,\star}$\\
\vspace{6pt}
$^{1}$State Key Laboratory for Mesoscopic Physics and Frontiers Science Center for Nano-optoelectronics, School of Physics, Peking University, Beijing 100871, China\\
$^{2}$School of Integrated Circuits and Electronics, MIIT Key Laboratory for Low-Dimensional Quantum Structure and Devices, Beijing Institute of Technology, Beijing, 100081, China\\
$^{3}$School of Physical Sciences, University of Chinese Academy of Sciences, Beijing 100049, China\\
$^{4}$Department of Physics, Renmin University of China, Beijing 100872, China\\
$^{5}$Collaborative Innovation Center of Quantum Matter, Beijing 100871, China\\
$^{6}$Yangtze Delta Institute of Optoelectronics, Peking University, Nantong 226010 Jiangsu, China\\
$^{7}$Liaoning Academy of Materials, Shenyang 110167, China\\
\vspace{3pt}
$^{\dag}$These authors contributed equally\\
$^{\star}$Corresponding to: yhuang@bit.edu.cn, wuzhou@ucas.ac.cn, tlxia@ruc.edu.cn, ye\_yu@pku.edu.cn
}

\begin{abstract}
\begin{adjustwidth}{-2cm}{0cm}
\textbf{The study of the mechanism of exchange bias phenomena and the achievement of its efficient control are of great importance, as it promotes the revelation of unique exchange interactions and the development of exotic applications. However, it is challenging due to the elusive interface between magnetic phases. In this study, we report an unprecedented exchange bias phenomenon observed in ultrathin uncompensated antiferromagnetic MnBi$_2$Te$_4$. The magnitude and direction of the exchange field can be intentionally controlled by designing a magnetic field sweep protocol without a field cooling process. The combined experimental and theoretical simulation results indicate that the spin-flip process assisted by the ubiquitous defect-induced pinning domain sites with varying inner exchange interactions might give rise to the emergence and robustness of this peculiar exchange bias. The temperature and thickness dependence of the exchange bias phenomena are systematically investigated for further study and exploitation of its unique properties. This mechanism hold promise for highly tunable exchange bias in prevalent magnetic systems by engineering the properties of domain structures, and also offers promising avenues for the design of spintronic devices combing its topology  based on MnBi$_2$Te$_4$.}
\end{adjustwidth}
\end{abstract}
\date{\today}

\maketitle

Since its discovery by Meiklejohn and Bean in 1956\cite{EB-1}, exchange bias (EB) has captivated researchers for over six decades, due to its potential applications in ultra-high density magnetic storage devices\cite{EBspinvalve} and other exotic applications\cite{EB-electricalcontrol,EB-beatingPM,EB-random}. Despite its wide-ranging technological significance, the fundamental mechanism underlying EB remains enigmatic, largely due to the intricate interplay of coexisting order parameters amidst a backdrop of complex magnetic disorder\cite{EB-5,EB-3,EB-4,16-maniv2021exchange,48-hauet2006training,49-keller2002domain,13-lachman2020exchange,14-noah2022tunable}. It is noteworthy that epitaxial bilayer often exhibits diminished EB effects compared to their polycrystalline counterparts, highlighting the pivotal role of defects in shaping this phenomenon\cite{18-takano1997interfacial}. The presence of disorder and/or defects at the interfaces facilitates the formation of domain structures, thus influencing the spin arrangements and exchange interactions between the ferromagnetic (FM) and antiferromagnetic (AFM) phases\cite{17-koon1997calculations,18-takano1997interfacial,20-ohldag2003correlation}. The interfaces associated with the domain and related domain wall (DW) are of paramount importance in biasing the entire FM phase\cite{48-hauet2006training,49-keller2002domain}. Therefore, the interplay between defects, domains, and magnetic reversal presents a novel avenue with the potential to yield groundbreaking giant EB phases across diverse systems. Nevertheless, navigating the interleaving energy landscape of multidomain structures presents a formidable challenge, as identifying energy barriers becomes a numerically challenging task\cite{li2004thermally}.

Exploration of the subtle and complex defect structures in magnetic materials is crucial because they often arise from competing internal forces at the atomic scale, which can be easily manipulated by external factors such as composition, temperature, or magnetic field to induce extraordinary phenomena. The recently discovered van der Waals (vdW) magnetic topological material, MnBi$_2$Te$_4$\cite{21-otrokov2019prediction,23-deng2020quantum,24-liu2020robust,27-deng2021high}, has emerged as a promising candidate for such investigations. The abundant defects present in MnBi$_2$Te$_4$, such as antisite substitutions and vacancies, significantly affect the magnetism of the material\cite{39-riberolles2021evolution,40-liu2021site,41-xu2022ferromagnetic,42-lai2021defect,43-yang2021odd,32-yan2019crystal,33-hou2020te,35-huang2020native,36-du2021tuning}. The rich interplay of defects and magnetic structures provides a unique platform to study EB phenomena with distinctive properties in the two-dimensional (2D) limit, opening a previously unexplored realm of microscopic exchange interactions.

\begin{figure*}[!tb]
    \includegraphics[width=1.5\columnwidth]{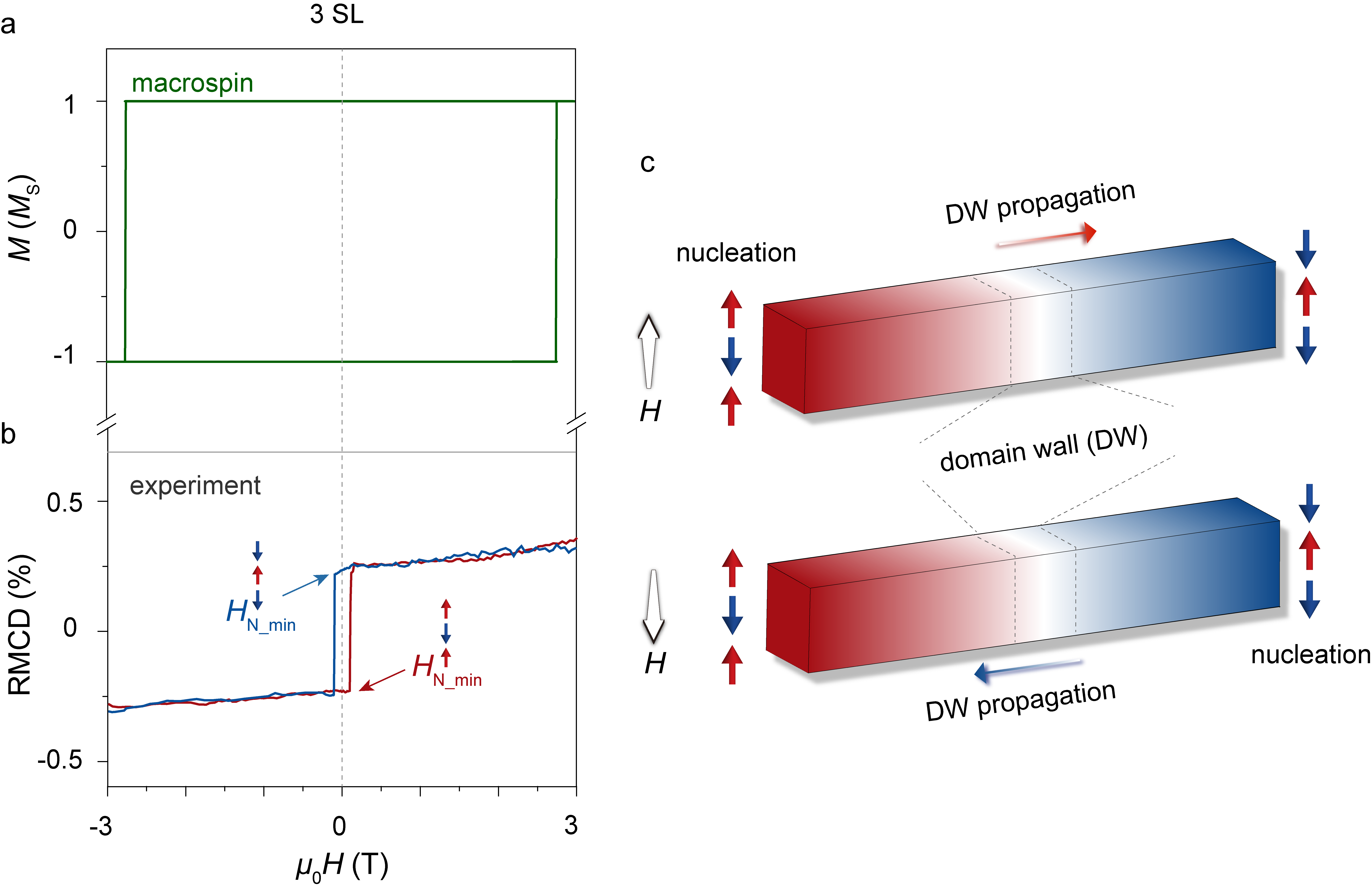}
    \captionsetup{singlelinecheck=off, justification = RaggedRight}
    \caption{\label{F1}\textbf{Magnetic reversal in odd-SL-numbered $\bf{MnBi_2Te_4}$}.
     \textbf{a-b.} Magnetic hysteresis loop of 3-SL MnBi$_2$Te$_4$ predicted by linear chain model (a) and measured by RMCD experiments (b) at low field.
    \textbf{c.} Schematic of the magnetic reversal process of the material with a domain nucleation field $H_{\rm{N}}$ much larger than the DW propagation field $H_{\rm{P}}$. The magnetic reversal is achieved by reverse domain nucleation and thus DW motion. 
}
\end{figure*}

Here, we reported a never-observed EB phenomenon in uncompensated AFM $\rm{MnBi_2Te_4}$ induced by defect-assisted magnetic reversal. In contrast to the previously reported EB phenomenon, where the bias field $H_{\rm{E}}$ depends on the field cooling (FC) process, both the direction and magnitude of $H_{\rm{E}}$ in the odd-number septuple layer (SL) MnBi$_2$Te$_4$ can be directly controlled by the field sweep protocol (defined by set field $H_{\rm{S}}$) at isothermal temperatures. Based on multi-domain theoretical simulations with rich comparative experiments, we confirm that this unique EB phenomenon originates from the evolution of the defect-related domain nucleation sites, which determine the magnetic reversal, with external $H_{\rm{S}}$. By revealing the complex exchange interactions around the defect regions, and its decisive influence on the domain nucleation which further induce the EB phenomenon, our findings not only pave the way for elucidating the underlying mechanisms that drive the magnetic behavior, but also promise to create innovative magnetic materials with tailored properties.

\begin{figure*}[!tb]
    \includegraphics[width=2\columnwidth]{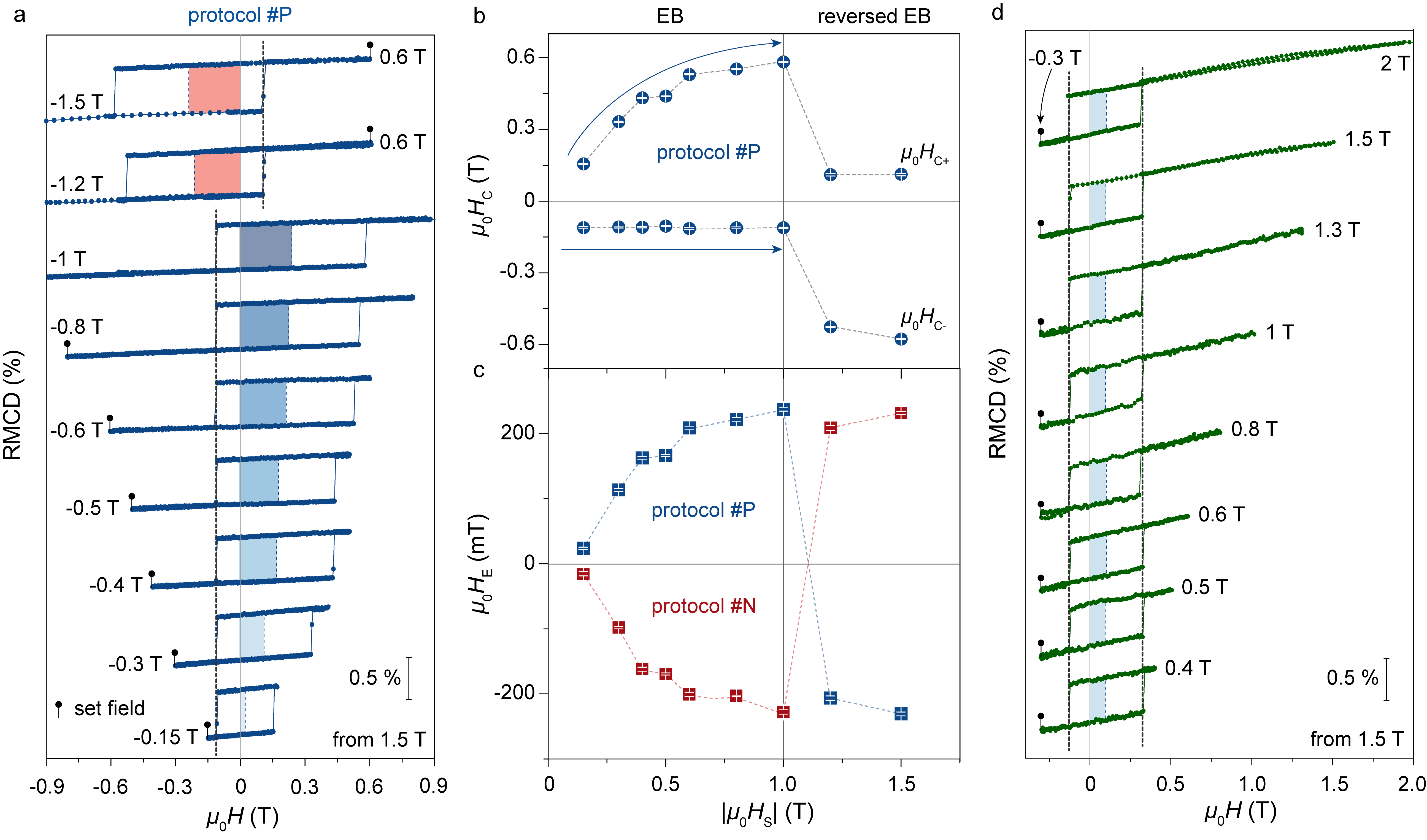}
    \captionsetup{singlelinecheck=off, justification = RaggedRight}
    \caption{\label{F2}\textbf{Exchange bias phenomenon of 5-SL $\bf{MnBi_2Te_4}$ at 2 K.} \textbf{a}. RMCD signal \textit{versus} magnetic field for the 5-SL sample at 2 K with a field sweep protocol {\footnotesize\#}P initialized at 1.5 T. With increasing negative set field, $\mu_0H_{\rm{C+}}$ increases from 0.15 T to 0.58 T, but $\mu_0H_{\rm{C-}}$ remains almost constant at $–$0.12 T. The exchange bias is reversed when the set field is greater than $-$1.2 T. \textbf{b}. The spin-flip fields $H_{\rm{C+}}$ and $H_{\rm{C-}}$ extracted from the hysteresis loops in \textbf{a} as a function of the set field $\left|\mu_0H_{\rm{S}}\right|$. The EB is divided by the set field into two regions, the increasing EB and inverted EB phases. \textbf{c}. The exchange field extracted by $H_{\rm{E}}$ = ($H_{\rm{C-}}$ + $H_{\rm{C+}}$)/2 under the protocol {\footnotesize\#}P (blue squares) and protocol {\footnotesize\#}N (red squares). \textbf{d} The consistent EB phenomenon when the set field $\mu_0H_{\rm{S}}$=$-$0.3 T is kept constant and the range of the opposite sweep field is increased.}
\end{figure*}

\bigskip
\noindent
\textbf{Spin flipping of odd-SL-numbered $\bf{MnBi_2Te_4}$}\\ 
\noindent
We first characterized the layer number-dependent magnetism of 1-SL to 5-SL MnBi$_2$Te$_4$ using reflective magnetic circular dichroism (RMCD) spectroscopy (Supplementary Fig. S1). Due to the A-type AFM nature, the 1-SL, 3-SL, and 5-SL MnBi$_2$Te$_4$ samples exhibited uncompensated magnetic hysteresis loops similar to those of ferromagnets near zero field, which is the focus of this study. It is interesting to find that the experimentally measured coercive field ($\sim$0.11 T) in the 3-SL sample was an order of magnitude smaller than the theoretical prediction ($\sim$2.76 T) from a linear chain model that considers the spins within the SL as ``macrospin" (Fig. \ref{F1}a-b). Taking into account the magnetic anisotropic energy and the interlayer exchange interaction of MnBi$_2$Te$_4$, the DW propagation field, $H_{\rm{P}}$, was significantly smaller than the reverse domain nucleation field, $H_{\rm{N}}$ (Supplementary Fig. S2). This indicates that once a reverse domain structure forms (i.e., domain nucleation), the DW will move rapidly until the magnetization of the entire sample reverse \cite{45-givord1988angular,46-franken2011domain,47-cullity2011introduction} as shown in Fig. \ref{F1}c, consistent with the observed rectangular hysteresis loops (Fig. \ref{F1}b). Note that there may exist multiple domain nucleation sites with varying $H_{\rm{N}}$, where the spin-flip field $H_{\rm{C}}$ coincides with the lowest nucleation field, i.e., $H_{\rm{C}}=H_{\rm{N\_{min}}}$.

\bigskip
\noindent
\textbf{Unique exchange bias phenomena}\\
Magnetic reversal of odd-number-layered MnBi$_2$Te$_4$ is dominated by the domain nucleation, and the influence of the possibly existed defects on the domain nucleation may lead to unprecedented EB effects. Therefore, we focus on demonstrating the unique EB phenomenon of 5-SL MnBi$_2$Te$_4$ under specific out-of-plane magnetic field sweep protocols. Similar EB behaviors were observed in other odd-number-layered samples (see Supplementary Note III). Under positive FC or initialization in a large positive field (e.g., 1.5 T) at 2 K, referred to as protocol {\footnotesize\#}P, the magnetic hysteresis loop of 5-SL MnBi$_2$Te$_4$ under a small sweep field depends strongly on the magnitude of the opposite magnetic field range (negative field here, defined as the set field $H_{\rm{S}}$). As $\mu_0H_{\rm{S}}$ increases from $-$0.15 T to $-$1.0 T, $\mu_0H_{\rm{C-}}$ remains nearly constant at $-$0.12 T, while $\mu_0H_{\rm{C+}}$ gradually increases from 0.15 T to 0.58 T (Fig. \ref{F2}a-b), resulting in a positive exchange field $\mu_0H_{\rm{E}}$ increasing from 22 mT to 235 mT as shown in Fig. \ref{F2}c. Comparably, under negative FC or initialization in a large negative field ($-$1.5 T), referred to as protocol {\footnotesize\#}N, as $\mu_0H_{\rm{S}}$ increases from 0.15 T to 1.0 T, $\mu_0H_{\rm{C+}}$ remains 0.12 T, while $\mu_0H_{\rm{C-}}$ gradually increases from $-$0.15 T to $-$0.57 T (Supplementary Fig. S3), leading to a negative exchange field $\mu_0H_{\rm{E}}$ from $-$17 mT to $-$230 mT. The magnitude of the EB is determined solely by $H_{\rm{S}}$, independent of the sweep range of the reversing field (Fig. \ref{F2}d and Supplementary Fig. S4). If the set field $\lvert H_{\rm{S}}\rvert \geq$1.2 T, the direction of the EB is reversed. For example, in protocol {\footnotesize\#}P, sweeping the field between $-$1.2 T ($\mu_0H_{\rm{S}}$=$-$1.2 T) and 0.6 T results in a $\mu_0H_{\rm{E}}$ of $-$207 mT (Fig. \ref{F2}a), which is consistent with the $\mu_0H_{\rm{E}}$ value of $-$203 mT obtained at $\mu_0H_{\rm{S}}$ of 0.6 T in protocol {\footnotesize\#}N. This indicates that $\lvert \mu_0H_{\rm{S}}\rvert$ of 1.2 T is sufficient to initialize the direction of EB. 

Encouragingly, by designing specific isothermal sweep protocols, the direction and magnitude of this unique EB can be easily adjusted without the need of warming and FC processes. Importantly, within a fixed-field sweep process, the EB field remains quite stable over multiple consecutive sweeps, showing no changes in $H_{\rm{C-}}$ and $H_{\rm{C+}}$ (see more in Supplementary Fig. S5-S6). This previously unobserved EB phenomenon is significantly different from the typical FM/AFM EB systems, which usually exhibit a simultaneous enhancement of both sides of $H_{\rm{C}}$ and an overall shift of the magnetic hysteresis loop\cite{EB-6}.

\begin{figure*}[!tb]    
    \includegraphics[width=2\columnwidth]{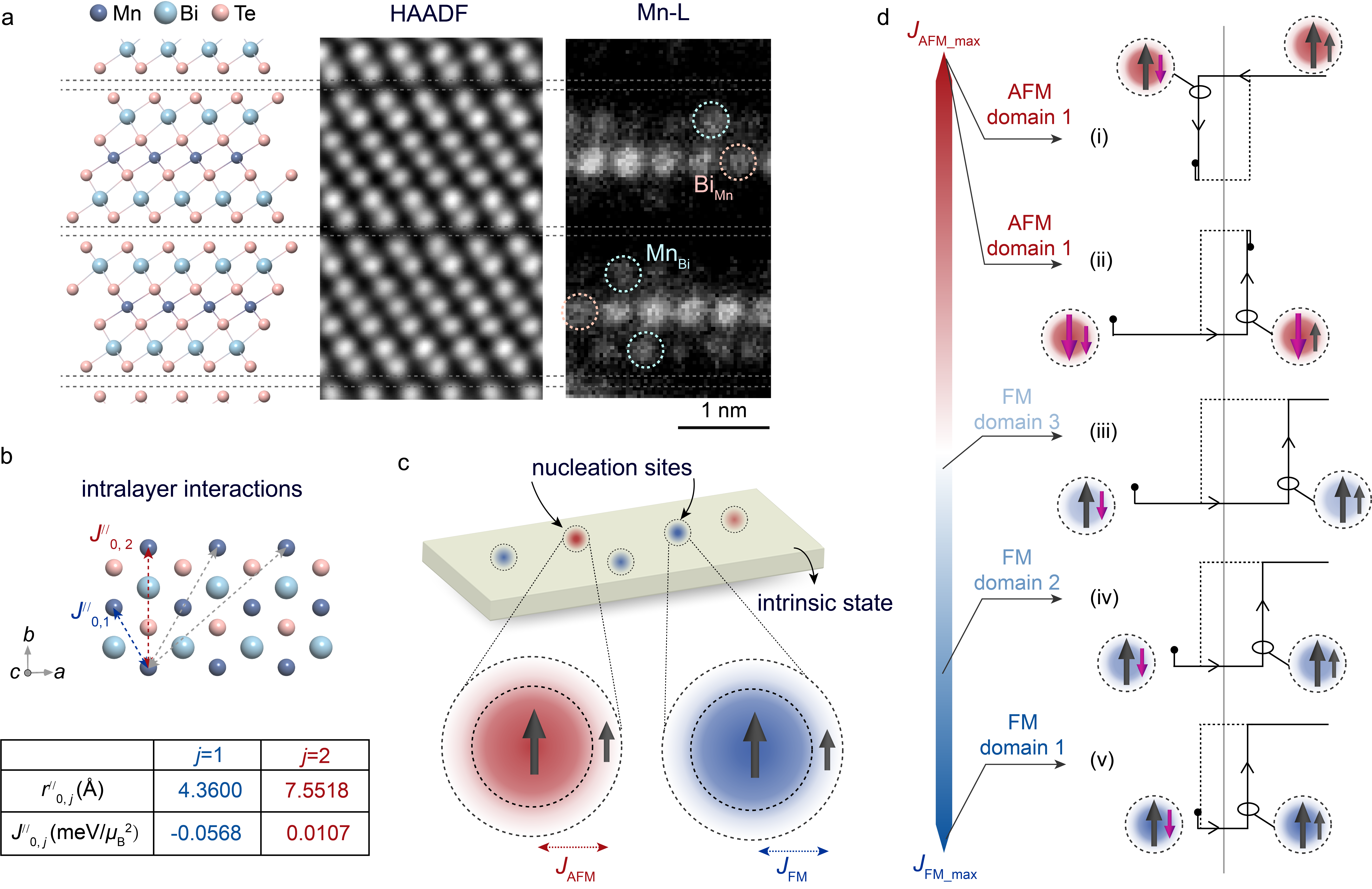}
    \captionsetup{singlelinecheck=off, justification = RaggedRight}
   \caption{\label{F3}\textbf{Phenomenological picture of exchange bias induced by defect-assisted magnetic reversal.}
    \textbf{a.} Cross-sectional atomic resolution HAADF-STEM image of MnBi$_2$Te$_4$ from the [100] direction and the corresponding EELS mapping of the Mn-L edge. The dashed circles indicate the Mn$_{\rm{Bi}}$ and Bi$_{\rm{Mn}}$ antisites, respectively. 
    \textbf{b.} First-principles calculations of the results of the strength of intralayer coupling between nearest (FM) and next nearest (AFM) neighboring Mn atoms.
    \textbf{c.} Schematic representation of the magnetic domain structure associated with randomly distributed defect regions in the sample, indicated by two black arrows of different sizes for the core and transition regions. The red (blue) colored background of the circles indicate that each domain is AFM (FM) coupled between the core and transition regions. 
    \textbf{d.} Schematic illustration of the spin flips triggered by domain-specific nucleation in the protocol {\footnotesize\#}P. The color gradient from red to blue represents the gradual transition from the strongest AFM coupling to the strongest FM coupling between the core region and transition regions at the different domain nucleation sites.}
\end{figure*}

\begin{figure*}[tb]
    \includegraphics[width=1.7\columnwidth]{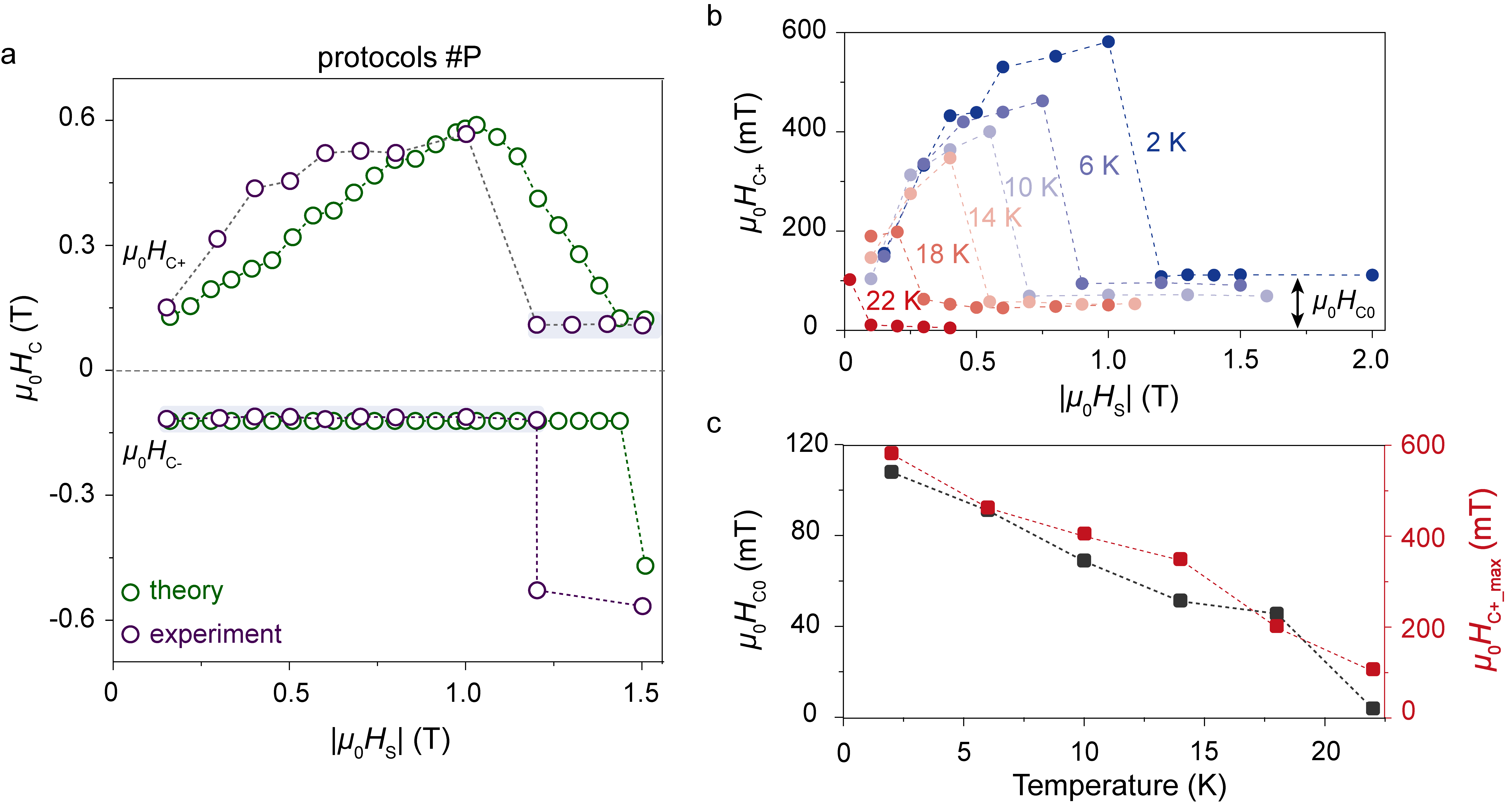}
    \captionsetup{singlelinecheck=off, justification = RaggedRight}
    \caption{\label{F4}\textbf{Theoretical calculations and temperature-dependent characterization of EB.}
    \textbf{a.} Theoretical calculations (green circles) and experimental results (purple circles) of $H_{\rm{S}}$ dependence of $H_{\rm{C+}}$ and $H_{\rm{C-}}$ for protocol {\footnotesize\#}P at 2 K.
    \textbf{b.} $H_{\rm{C+}}$ \textit{versus} $H_{\rm{S}}$ for 5-SL sample as temperature varies.
    \textbf{c.} Extracted $H_{\rm{C0}}$ and $H_{\rm{C+}\_max}$ with increasing temperature.}
\end{figure*}

\bigskip
\noindent
\textbf{Phenomenological picture of defect-assisted spin flipping}\\ 
The enhancement and nonmonotonic variation of $H_{\rm{C+}}$ (take protocol {\footnotesize\#}P as an example) and resulting $H_{\rm{E}}$ in 5-SL MnBi$_2$Te$_4$ suggest that the sweeping set field may induce magnetic reversal dominated by different domain nucleation sites with different $H_{\rm{N}}$. Notably, in the even-number-layered MnBi$_2$Te$_4$, the magnetic hysteresis loop induced by interfacial effects\cite{43-yang2021odd} does not show any EB effects (Supplementary Note IV), indicating that the domain nucleation sites governing the magnetic reversal are internal to the material. To investigate the origin of these domain nucleation sites, single crystal X-ray diffraction experiments were first performed (Supplementary Table 1). The MnBi$_2$Te$_4$ sample used was determined to have a chemical stoichiometry of Mn$_{0.86}$Bi$_{2.14}$Te$_4$, revealing 3\% $\rm{Mn_{Bi}}$ and 20\% $\rm{Bi_{Mn}}$ antisite defects. This was further validated by an atomic-resolution high-angle annular dark-field (HAADF) cross-sectional STEM image of MnBi$_2$Te$_4$ taken along the [100] direction and the corresponding electron energy loss
spectroscopy (EELS) mapping (Fig. \ref{F3}a), showing a weakening of the Mn signal in the Mn layer and the presence of Mn signals in the Bi layer (see more in Supplementary Fig. S10). First-principles calculations (Fig. \ref{F3}b) indicate that the nearest neighbors (blue dashed line) and next nearest neighbors (red dashed line) of Mn atoms within the layer exhibit FM ($J_{\rm{FM}}$) to AFM ($J_{\rm{AFM}}$) couplings, with the coupling strength decreasing by a factor of 5. This is in agreement with previous results\cite{21-otrokov2019prediction}. Due to the prevalent distribution of defects, some regions may exhibit elusive exchange interactions, possibly ranging from FM to AFM coupling. By randomly distributing Mn atoms in a 100$\times$100 array with 20\% vacancy rate, we identified the existence of Mn atoms which only possess the next nearest AFM coupled Mn neighbors (Supplementary Fig. S11). While most Mn atoms have the nearest FM coupled Mn neighbors, their FM coupling strength may also vary. The regions with local high defect concentrations are randomly distributed throughout the sample and can be very small, making them indistinguishable when using a $\sim$1 $\mu$m diameter laser spot for RMCD spatial maps (Supplementary Fig. S11).

Based on these experimental observations, we propose a physical picture of defect-assisted domain nucleation driving magnetic reversal that aims to capture all EB signatures in the MnBi$_2$Te$_4$ sample (see further discussion in Supplementary Note VI). First, we simplify the magnetic structures at the defect regions into two parts, i.e., a core pinning region with high defect concentration and a transition region with a decreased concentration as shown in Fig. \ref{F3}c. Then the exchange interactions inside these defect regions (that is, the coupling between the core and transition region) can be simplified by a two-spin model with averaged FM or AFM coupling. This averaged coupling strength also vary at different defect regions due to the different defect concentration or types, indicated by color changes from blue to red in Fig. \ref{F3}d. As discussed before, once a reverse domain structure is formed, it triggers spin flipping throughout the whole sample, so we only need to consider the domain structure evolutions in these defect regions.

The core and transition regions are represented by $\uparrow$ and $\downarrow$ arrows of different sizes in Fig. \ref{F3}d. Using protocol {\footnotesize\#}P as an example, after initialization with a 1.5 T magnetic field, the core and transition regions are in a $\uparrow$ state. When the magnetic field is decreased to a negative value, the transition region with the strongest $J_{\rm{AFM}}$ coupling to the pinned core region (AFM domain 1) first undergoes a downward flip (Fig. \ref{F3}d(i)). Once the transition region forms a reversed domain structure, the DWs will immediately propagate throughout the sample to achieve magnetic reversal, thus $H_{\rm{C-}}$ is determined by the flip field of AFM domain 1. Without considering the reversal of the pinned core region under $H_{\rm{C-}}$, as the field sweeps toward positive values, the transition region with the strongest $J_{\rm{FM}}$ coupling to the pinned core (FM domain 1) flips upward first (Fig. \ref{F3}d(v)), at this point, $H_{\rm{C+}}$ is determined by the flip field of FM domain 1. However, if we further sweep the negative field after $H_{\rm{C-}}$, i.e., increasing the set field $H_{\rm{S}}$, the pinned core region will gradually depin and flip downward. The pinned core region of FM domain 1 will flip first, while that of AFM domain 1 will flip last, since the transition regions and the entire intrinsic region have already flipped downward below $H_{\rm{C-}}$. As shown in Fig. \ref{F3}d(iv), with increasing $H_{\rm{S}}$, the pinned core region of FM domain 1 has depinned, leading to the magnetic reversal at $H_{\rm{C+}}$ determined by the transition region of FM domain 2, which has a lower FM coupling strength to the core region and larger flip field. When the pinned core of FM domain 2 depins under increasing $H_{\rm{S}}$, a larger $H_{\rm{C+}}$ is provided by FM domain 3 (Fig. \ref{F3}d(iii)), and so on. The depinning of the AFM domain 1 is the most difficult, thus spin flips during descending field always occur at the same $H_{\rm{C-}}$ that is determined by the flip field of AFM domain 1. Only when $H_{\rm{S}}$ exceeds the depinning field of AFM domain 1 do all pinned magnetic domain structures associated with defects flip simultaneously (Fig. \ref{F3}d(ii)). This is consistent with protocol {\footnotesize\#}N, which underwent a large negative field initialization. Our phenomenological model effectively reproduces the experimental results - the depinning of the domain structures at the defect regions leads to changes in nucleation sites during defect-assisted spin reversal, resulting in a unique exchange bias phenomenon dependent on $H_{\rm{S}}$.



\bigskip
\noindent
\textbf{Modeling of defect-assisted magnetic reversal}\\ 
\noindent
To quantify the consistency between the model and experimental results, we used the classical two-spin Heisenberg model to simulate the magnetic reversal evolution under field sweep protocols. The magnetizations in the core and transition regions are denoted $N_cM_S$ and $N_tM_S$, respectively, where $N$ represents the number of spins in each region. Assuming that two spins line up in the same plane, the energy can be written as

\begin{multline}\label{E1}
E=\mu_0M_{\rm{S}}N_c[-H(\mathrm{cos}\phi_{ci}+L\mathrm{cos}\phi_{ti})+\frac{H_{Kc}}{2}\mathrm{sin}^2\phi_{ci}\\ 
+\frac{LH_{Kt}}{2}\mathrm{sin}^2\phi_{ti}+\frac{LH_{J}}{2}\mathrm{cos}(\phi_{ci}-\phi_{ti})]
\end{multline}

\noindent
Here, $L$ is equal to $N_c/N_t$, and the anisotropy energies of the core and transition regions are scaled by the magnetic field, i.e., $H_{Kc}(H_{Kt}) = K_c(K_t)/\mu_0M_S$, where $H_J$ reflects the averaged interaction between the transition and core regions, ranging from negative (FM coupling) to positive (AFM coupling). Considering various pinning regions with different $L$ and $H_J$ values in the sample, we proposed a system with defects satisfying $\mu_0H_{Kc}$ = 1 T, $\mu_0H_{Kt}$ = 0.6 T, 2 $\le L \le$ 2.7, and $\left|\mu_0H_J\right|$$\le$ 0.44 T (see more discussions in Supplementary Note VI). Based on the experimental field sweep range, we simulated the spin reversal field under protocol {\footnotesize\#}P with $\mu_0H_{\rm{S}}$ ranging from $-$0.16 T to $-$1.5 T using the model (simulation details are discussed in Methods). As shown in Fig. \ref{F4}a, the model extracted the general trend of the spin reversal, and the magnitude of the reversal field is in good agreement with the experimental results. Minor discrepancies between the model and experimental results may come from the simplified parameter boundaries of $L$ and $H_J$ of each pinning site. The simulation results show a distinct continuous decrease in $H_{\rm{C+}}$ as $\mu_0H_{\rm{S}}$ increases from $-$1 T to $-$1.5 T under protocol {\footnotesize\#}P, while the experimental $H_{\rm{C+}}$ immediately decreases to a minimum and the EB direction reverses when $H_{\rm{S}}$ reaches 1.2 T. From the perspective of the defect-assisted spin reversal picture discussed above, this suggest that pinning regions with AFM coupling undergo reversal within a narrow field range (between 1 T and 1.2 T).

\bigskip
\noindent
\textbf{Temperature dependence of EB}\\
Finally, we examine the temperature-dependent characteristics of the EB phenomenon under field sweep protocols in the 5-SL MnBi$_2$Te$_4$ (see 3-SL sample results in Supplementary Note VII). Taking protocol {\footnotesize\#}P as an example, this unique EB phenomenon persists with increasing temperature to $T_{\rm{N}}$, where the $H_{\rm{E}}$ increase with increasing $\left|H_{\rm{S}}\right|$ and finally reverse its direction at $\left|H_{\rm{S0}}\right|$ (Fig. \ref{F4}b). $\left|H_{\rm{S0}}\right|$ represents the depinning field of the defect region with the maximum AFM coupling (AFM domain 1). Due to the increasing thermal flunctuations, $\left|H_{\rm{S0}}\right|$ and the maximum achievable $H_{\rm{C+}}$ ($H_{\rm{C+\_max}}$) decrease with increasing temperature. Similarly, the value of $H_{\rm{C+}}$ after EB direction reversal (defined here as $H_{\rm{C0}}$), which represents the coupling strength between the core region and transition regions in AFM domain 1 also decreases with increasing temperature (Fig. \ref{F4}c). It should be noted that this unique EB phenomenon disappears at a temperature above 22 K as the hysteresis loop vanishes in MnBi$_2$Te$_4$ (Supplementary Fig. S16), providing theoretical basis for achieving EB phenomena with higher tempereture (limited by the intrinsic critical temperature of magnetic materials but not the $T_{\rm{N}}$ of AFM/FM systems).

\bigskip
\noindent
\textbf{Conclusion}\\
In conclusion, our study has revealed the unique EB phenomena achieved in atomically thin MnBi$_2$Te$_4$, providing theoretical evidence for defect-assisted magnetic reversal. By simply controlling the isothermal magnetic field sweep protocol, unprecedented adjustments in the magnitude and direction of the EB field can be achieved. We conclude that this distinctive EB phenomenon arises from the various pinned domain structures induced by prevelant defect regions in MnBi$_2$Te$_4$ sample. The $H_{\rm{S}}$-dependent magnetic domain nucleation site determines the evolution of spin flip field $H_{\rm{C}}$. Under appropriate magnetic anisotropy and exchange coupling energies, the effective creation and control of magnetic domain formation and properties will allow this EB mechanism to be observed in extended magnetic material systems. We note that very recently, several groups have also found shifted magnetic hysteresis loop in intrinsic MnBi$_2$Te$_4$ or MnBi$_2$Te$_4$ heterostructures\cite{124EB-WKL,124EB-CCZ,124CGTEB,124CIEB-2}. While our phenomenological model combined with simulations can largely track the experimental observations, further elucidation of the domain structure evolution is necessary. For example, rigorous atomic-scale investigations using single-spin sensitive magnetic measurement techniques\cite{52-song2021direct} are essential to probe the defect-induced pinning sites and their domain distribution during magnetic reversal. In addition, tuning the defect density and species may provide opportunities for tailored EB with significant potential applications in spintronic devices.

\bigskip

\bibliography{ref}

\clearpage
\newpage

\noindent
\textbf{Methods}\\
\noindent
\textbf{Growth of MnBi$_2$Te$_4$ crystals}.
MnBi$_2$Te$_4$ single crystals were fabricated by the self-flux method. The mixture was loaded into a corundum crucible in the ratio of Mn: Bi: Te = 1: 8: 13 (MnTe: $\rm{Bi_2Te_3}$ = 1: 4) and sealed in a quartz tube. The quartz tube was then placed in a furnace and heated to 1000 $^{\circ}$C for 20 hours for sufficient homogenization. The mixture was cooled rapidly at 5 $^{\circ}$C/h to 605 $^{\circ}$C and then slowly at 0.5 $^{\circ}$C/h to 590 $^{\circ}$C and held for 2 days. Finally, bulk single crystals were obtained by centrifuging.
\bigskip

\noindent 
\textbf{RMCD measurements}. RMCD measurements were performed on an Attocube closed-cycle cryostat (attoDRY2100) with temperatures down to 1.6 K and out-of-plane magnetic fields up to 9 T. The 633 nm HeNe laser was modulated between left and right circular polarization by a photoelastic modulator (PEM) and focused onto the sample through a high-numerical-aperture (0.82) objective. The reflected light was detected by a photomultiplier tube (Thorlabs PMT 1001/M). Magnetization under the external magnetic field was detected by the RMCD signal, which was determined by the ratio of the PEM-modulated 50.052 kHz a.c. signal to the chopper-modulated 789 Hz a.c. signal (processed by a  Zurich HF2LI two-channel lock-in amplifier). 
\bigskip

\noindent 
\textbf{STEM characterization}. Samples for cross-sectional investigations were prepared by standard lift-out procedures using a focused ion beam system. Aberration corrected STEM imaging was performed using a Nion HERMES-100 operated at 60 kV. The probe forming semi-angle was set to 32 mrad. HAADF images were acquired using an annular detector with a collection semi-angle of 75–210 mrad. EELS measurements were performed using a collection semi-angle of 75 mrad, an energy dispersion of 0.3 eV per channel, and a probe current of ~20 pA. The Te-M (572 eV) and Mn-L (640 eV) absorption edges were integrated for elemental mapping after background subtraction. The original spectrum images were processed using principal component analysis (PCA) to reduce random noise.

\bigskip
\noindent 
\textbf{Theoretical modeling of multi-domains}. In our simulations, the magnetic field starts with a positive magnetic field $H_{\rm{start}}$ that is large enough to ensure that all spins are polarized to one direction, and then we gradually sweep the field to $H_{\rm{S}}$ (in the case of the negative set fields) and finally back to $H_{\rm{start}}$. At each magnetic field, all defect regions are first evolved independently from the final state of the previous magnetic field using the gradient descent method. Here the step size of each iteration is less than 1×10$^{-3}$ rad, the initial $\theta_{i}$ is less than 1×10$^{-5}$ rad, and the condition for convergence is that the number of iterations is larger than 10,000 and $\left|\frac{1}{M_{\rm{S}}N_c}\frac{\partial E}{\partial\theta_i}\right|$ is less than 1×10$^{-9}$. After that, if we find one or more spin flips in the transition region, which implies the formation of a reverse domain structure, then the overall intrinsic state is flipped immediately due to the DW motion. The simulated nucleation fields under several $H_{\rm{S}}$ are shown in the main text, and the detailed illustration are presented in Supplementary Note VI. However, there are still discrepancies between the simulation results of the model and the experimental results. First, the predicted $H_{\rm{C+}}$ of our model in the EB phase is slightly smaller than the experimental value. One possible explanation is that the two-spin model is oversimplified and loses some of the fine structure and behavior of the real defect region. Second, from our simulation results, there is a continuous downward trend of $H_{\rm{C+}}$ between the EB and the reverse EB phases, whereas in the experimental observations there is an abrupt reversal at $-$1.2 T. We attribute this to the small number of AFM-coupled domain sites and possible thermal fluctuations affecting the reversal of some defects-induced pinning sites. We also note that although the chosen parameters give results in good agreement with the experimental observations, the theoretical and experimental evidence on the decreased $K_c$ or $K_t$ and the range of $L$ and $H_{\rm{J}}$ distributions needs to be further confirmed and optimized.

\bigskip
\noindent
\textbf{Data availability}\\
The data that support the findings of this study are available from the corresponding author upon reasonable request.

\bigskip
\noindent
\textbf{Acknowledgement}\\
This work was supported by the National Key R\&D Program of China (No. 2022YFA1203902), the National Natural Science Foundation of China (No. 12250007 and No. 62274010), Beijing Natural Science Foundation (No. JQ21018), the China Postdoctoral Science Foundation (2023TQ0003 and 2023M740122), the Young Elite Scientists Sponsorship Program by CAST(2023QNRC001), and CAS Project for Young Scientists in Basic Research (YSBR-003). This research benefited from resources and supports from the Electron Microscopy Center at the University of Chinese Academy of Sciences.

\bigskip
\noindent
\textbf{Author contributions}\\
S.Y., X.X., and Y.Y. conceived the project. S.Y. and X.X. performed the measurements and data analysis. Y.G. carried out the theoretical simulations. P.G. performed the ab initio calculations. H.W. and T.X. grew the MnBi$_2$Te$_4$ bulk crystals and Y.H. prepared the atomically thin samples. R.G. performed the STEM measurements under the supervision of W.Z.. S.Y., X.X., and Y.Y. wrote the paper with contributions from other authors.

\bigskip
\noindent
\textbf{Competing interests}\\
The authors declare no competing interests.

\end{document}